\documentclass[showpacs,twocolumn,secnumarabic,amssymb, nobibnotes, aps,
prb]{revtex4-1}
\usepackage{graphicx} 
\usepackage{times}

\def\U#1{{%
\def\O{\mbox{O}}
\def\u{\mbox{u}}
\mathcode`\u=\mu
\mathcode`\O=\Omega
\mathrm{#1}}}
\def\sub#1{_{\rm #1}}

\begin{document}

\title{
Storage of electromagnetic waves in a metamaterial that mimics
electromagnetically induced transparency}

\author{Toshihiro Nakanishi, Takehiro Otani, Yasuhiro Tamayama, and Masao Kitano}%
%\email[E-mail: ]{t-naka@kuee.kyoto-u.ac.jp}
\affiliation{Department of Electronic Science and Engineering, Kyoto University, Kyoto 615-8510, Japan}
\date{15 April 2013}%

\begin{abstract}
 We propose a method for dynamically
 controlling the properties of a metamaterial that mimics
 electromagnetically induced transparency (EIT) 
 by introducing varactor diodes to manipulate the structural symmetry
 of the metamaterial. 
 Dynamic modulation of the EIT property enables 
 the storage and retrieval of electromagnetic waves. 
We confirmed that the electromagnetic waves  were stored and released,
while maintaining the phase distribution in the propagating direction.
\end{abstract}


\maketitle

The group velocity of electromagnetic waves, which 
represents the propagation speed of the pulse envelope, can be modified in 
dispersive media.
Electromagnetically induced transparency (EIT),
which realizes  highly dispersive media and leads to a slower light propagation speed,
was originally investigated in atomic physics.
\cite{Harris1997,Fleischhauer2005}
The EIT effect enables the propagation of light
through an initially opaque medium by the incidence of auxiliary light,
which is called control light. 
Furthermore,
the group velocity of the light is dramatically decreased
in the transparent region. \cite{Hau1999}
Studies of the slowing of light using EIT phenomena have developed
the storage of optical pulses. \cite{Fleischhauer2000}
The storage of light 
can be realized by dynamically reducing the group velocity 
when the light pulse is trapped in an EIT medium.
The storage of light using EIT is a promising 
technology, which leads to the realization of optical memory and quantum memory.
\cite{Phillips2001, Liu2001, Turukhin2002}

EIT effects are not unique to atomic systems.
Similar effects have been observed in
various systems, such as optical waveguides
equipped with cavities \cite{Little1999,Xu2006,Totsuka2007,Kekatpure2010}
and circuit models. \cite{GarridoAlzar2002,Nakanishi2005}
In recent years, the implementation of EIT effects in metamaterials
has attracted considerable attention.
The metamaterial is an assembly of subwavelength structures,
which are designed to realize unconventional electromagnetic properties.
It is possible to design metamaterials that mimic EIT effects,
and these metamaterials are sometimes referred to as EIT metamaterials.
\cite{Zhang2008a,Tassin2009,Luk'yanchuk2010}
Each element in an EIT metamaterial has  two resonant modes,
one of which is a highly radiative mode coupled to propagating waves, 
which has a low quality factor.
The other is a ``trapped mode'' with a high quality factor,
which is excited only through coupling with the radiative mode,
not by the propagating waves directly.
If the resonant frequency $\omega\sub{T}$ of the trapped mode is 
tuned to around that of the radiative mode,
and if the incident waves are oscillating at $\omega\sub{T}$,
the energy excited in the radiative mode is transferred to
the trapped mode with low-energy dissipation.
As a consequence, a suppression of the dissipation 
in the metamaterial, or the EIT effect, is observed.
The physical origin of the transparency in the metamaterial 
differs from that of an EIT medium composed of three-level atoms with two 
ground states and a common excited state, but there are good
 correspondences between the two systems. The coherent oscillation
 between a ground state and the excited state 
in an atomic EIT system corresponds to 
the resonance in the radiative mode in EIT metamaterial, 
and the coherent oscillation between two ground levels 
with long relaxation time corresponds to the resonance in the trapped mode.
EIT metamaterials can be implemented in various frequency regions
by applying the same design rules,
and a variety of EIT metamaterials have been demonstrated in 
the microwave, \cite{Fedotov2007,Papasimakis2008,Kurter2011,Kang2011} 
terahertz, \cite{Chiam2009,Li2011,Singh2011,Liu2012} and optical
regions. \cite{Liu2009,Zhang2010,Huang2011a}
In addition to the slow propagation of electromagnetic waves, 
applications for accurate sensing \cite{Dong2010}
and manipulation of the near field \cite{Zhang2012}
have also been investigated
by taking advantage of the sharp resonance.

The storage of electromagnetic waves  is one of the most fascinating applications
of the EIT effect.
To achieve this storage, the coupling between
the trapped mode and the radiative mode must be dynamically controlled.
In most EIT metamaterials,
control of the coupling has been realized 
by mechanically tuning the relative positions of two resonators, and
dynamic control is difficult.
Some researchers have succeeded in the semi-static control 
of EIT properties
by changing the incident angles \cite{Tamayama2012,Jin2012} or 
by structural modification induced by carrier excitation in a semiconductor.
\cite{Gu2012}
However, the storage and retrieval of electromagnetic waves
has not been reported.
The main goal of this Rapid Communication is to realize the storage and retrieval
of electromagnetic waves in a metamaterial.
Extending the idea of EIT metamaterials 
with broken structural symmetry, \cite{Fedotov2007,Kang2011,Singh2011}
we propose an EIT metamaterial with variable capacitors (varactors)
installed to dynamically control the structural symmetry,
which determines the coupling between two resonant modes.
We first describe the design and operating principles of the
metamaterial, while discussing simulation results.
We also show experimental results demonstrating semi-static control of the EIT
properties of the metamaterial in the microwave region.
In the latter part of this Rapid Communication,
we experimentally demonstrate 
the storage and retrieval of electromagnetic waves,
employing dynamic modulation of the EIT properties.
We also confirm that the waves are stored and released
while maintaining the phase distribution in the propagating direction.

\begin{figure}[t]
 \begin{center}
  \includegraphics[scale=1]{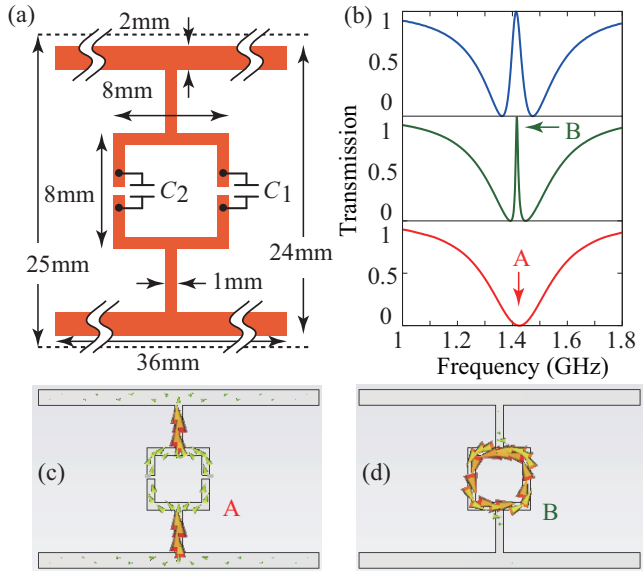}
  \caption{(color online) (a) Unit structure and its dimensions. (b)
  Transmission spectra for 
  $C_1=C_2=2.1 \U{pF}$ (bottom),  
  $C_1=1.91 \U{pF}, C_2=2.33\U{pF}$ (middle), 
  and $C_1=1.75 \U{pF}, C_2=2.63\U{pF}$ (top).
  Current distributions for (c) point {\rm A}  and (d) point {\rm B}.
}
  \label{fig1}
 \vspace{-3mm}
 \end{center}
\end{figure}

Figure \ref{fig1}(a) shows a unit cell of our metamaterial, which is made of metal. 
Two variable capacitors, $C_1$ and $C_2$, are introduced 
at two gaps in the center loop to dynamically control the EIT properties.
We calculate the transmission properties of the metamaterial
using an electromagnetic simulator ({\textsc{cst mw studio}}) 
for various pairs of values for $C_1$ and $C_2$.
In the simulation, periodic boundary conditions are set for
the unit cell with a size of $120\,\U{mm} \times 25\,\U{mm}$
to simulate an infinite array of the unit structures.
The incident waves are vertically polarized.
The results are shown in Fig.~\ref{fig1}(b) for
  $C_1=C_2=2.1 \U{pF}$ (bottom),  
  $C_1=1.91 \U{pF}, C_2=2.33\U{pF}$ (middle), 
  and $C_1=1.75 \U{pF}, C_2=2.63\U{pF}$ (top).
For $C_1=C_2$, the transmission is depressed over
a broad spectrum.
At the center of the depression,
a resonant current is induced on the metal, as shown
in Fig.~\ref{fig1}(c).
This mode can be regarded as a radiative mode.
On the other hand, for $C_1 \neq C_2$, 
sharp transparent regions appear in
the broad resonant line.
At the transmission peak,
a resonant loop current is induced, as shown 
in Fig.~\ref{fig1}(d), and this can be regarded as a trapped mode,
which has a much higher quality factor than the radiative mode  
and is uncoupled from incident wave.
(The structure is designed so that the resonant frequencies
of the two modes coincide.)
The electromagnetic energy received in the radiative mode is transferred
into the trapped mode through a coupling,
which is provided by breaking the symmetry of the metamaterial.
As a result, dissipation is suppressed and the metamaterial
is rendered transparent.
The increase in the degree of asymmetry results in 
the broadening of the transparent window,
as is clearly shown in the middle and top panels of Fig.~\ref{fig1}(b).
It is possible to control the width of the transparent window 
$\Delta \omega$ without changing the center of the window.
This is accomplished by keeping the composite capacitance in the loop,
$C\sub{L}=C_1 C_2 / (C_1+C_2)$, constant.
The minimum of $\Delta \omega$ is limited by the linewidth of an
isolated trapped mode,
which is determined by the radiation loss of the loop current  
and Ohmic loss in the capacitors.
The transparency peak is deteriorated by the loss in the trapped mode.

\begin{figure}[]
 \begin{center}
  \includegraphics[scale=1]{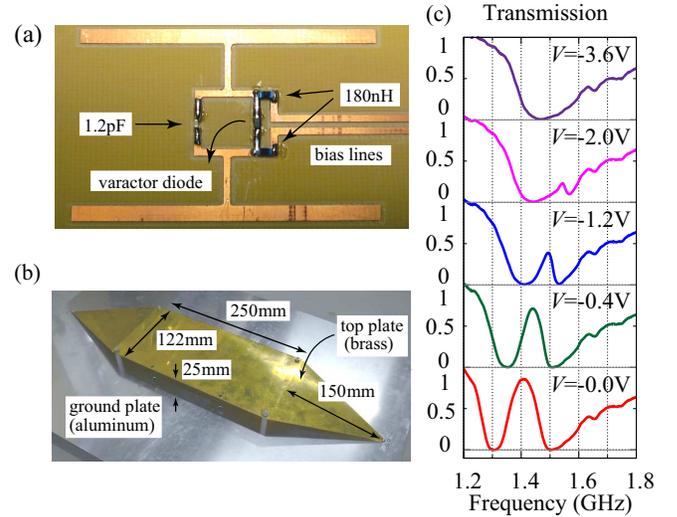}
  \caption{(color online) Photographs of (a) the unit structure and (b)
  the waveguide. (c) Transmission spectra for
  various bias voltages.}
  \label{fig2}
 \vspace{-3mm}
 \end{center}
\end{figure}

To experimentally demonstrate the EIT effects in the microwave region,
we fabricated the proposed structures using a copper film 
on a dielectric substrate with a permittivity of 3.3 and a thickness of
$0.8\,\U{mm}$, as shown in Fig.~\ref{fig2}(a).
In one side of the loop,
we introduced a varactor diode (Infineon BBY52-02W),
whose capacitance $C_1$ is a function of the applied bias voltage $V$.
The bias voltage was fed through $180$-$\U{nH}$ inductors,
which were introduced to pass only the bias voltage 
and to isolate the structure from the bias circuitry for signals at
the operating frequency of the metamaterial of over $1\,\U{GHz}$.
Ohmic loss in the varactor diode,
which is not negligible,
degrades the quality factor of the trapped mode.
Therefore, we used a normal capacitor ($C_2=1.2\,\U{pF}$) with low Ohmic loss
at the other side.
In this situation, unlike a simulation in which $C\sub{L}$ is kept constant,
the center and width of the transparent window
are expected to change with $C_1(V)$.
For measurements, we introduced a stripline waveguide
with tapered structures, as shown in Fig.~\ref{fig2}(b) (see details in
Ref.~\onlinecite{Nakanishi2012b}).
Three layers of the metamaterial, each of which contained a
single structure, were prepared.
We placed these layers in the waveguide with a separation of
$7\,\U{cm}$,
which was sufficient to avoid undesired coupling between the layers,
and measured the transmission through the metamaterial.
The results of these transmission measurements for various $V$
are shown in Fig.~\ref{fig2}(c).
For $V=-3.6\,\U{V}$,
a broad depression in the transmission is observed.
This indicates that the elements of the metamaterial are symmetric,
i.e., $C_1(V) \simeq C_2=1.2\,\U{pF}$, which agrees with the value in 
the varactor diode data sheet.
By breaking the symmetry, 
we can observe transparent windows.
When increasing $C_1(V)$ by reducing  $|V|$,
the asymmetry of the structure is enhanced, 
and the transparent window becomes wider.
The transmission peaks shift to a lower frequency
because they correspond to the resonant frequencies of the trapped modes,
which are determined by the loop capacitance 
$C\sub{L}$.

\begin{figure}[t]
 \begin{center}
  \includegraphics[scale=1]{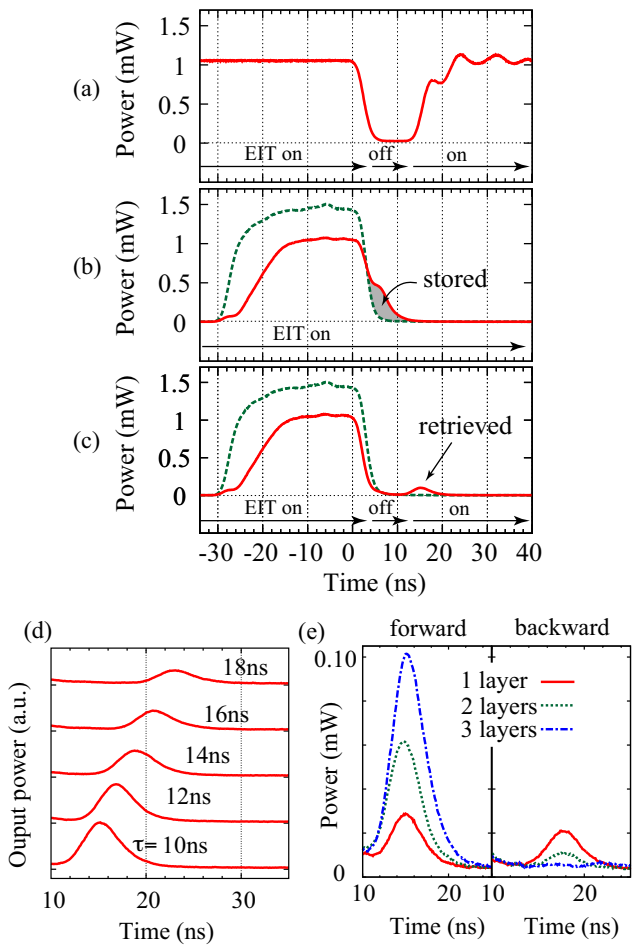}
  \caption{(color online) (a) Modulation of a continuous wave. 
  (b) Original pulse (dotted line) and
  transmitted pulse (solid line) for $V=-0.4\,\U{V}$. (c) 
  Storage and retrieval of electromagnetic energy for a control pulse
  with a width of $\tau=10\,\U{ns}$.
  (d) Retrieved pulses for various $\tau$. (e) Retrieved pulses
  propagating forward (left) and backward (right) for various $n$.}
  \label{fig3}
 \end{center}
\end{figure}

Next, we demonstrate the modulation of continuous waves by varying 
$V$ with time to confirm the dynamic control of the EIT effect
and to identify the timing and transient time of the modulation.
By using a pulse generator,
we apply negative rectangular pulses with an amplitude of $3.2\,\U{V}$,
an offset of $-0.4\,\U{V}$, and a width of $\tau=10\,\U{ns}$, 
at the same time for all of the varactors.
With this setting,
the transmission property of the metamaterial switches from 
the second lowest spectrum ($V=-0.4\,\U{V}$) to the top spectrum ($V=-3.6\,\U{V}$)   in
Fig.~\ref{fig2}(c).
A signal generator feeds continuous waves at $1.440\,\U{GHz}$,
which corresponds to the transmission peak for $V=-0.4\,\U{V}$,
into the waveguide containing the three-layered metamaterial.
We record the power of the waves transmitted through the waveguide.
Figure \ref{fig3}(a) shows transmitted microwave power 
under the modulation of the metamaterial.
The origin for time $t$ is set to the moment when the signal starts to decrease.
For $t<0$, the bias is $V=-0.4\,\U{V}$,
and the incident waves can travel through the metamaterial
because of the EIT effect.
After some transient time of around $4\,\U{ns}$ from $t=0$,
almost all of the energy is blocked by the metamaterial,
because the EIT effect disappears for $V=-3.6\,\U{V}$.
At $t=\tau+t\sub{T}$, where $t\sub{T} (\sim 3\,\U{ns} )$ is the transient
time for 
a transparent window to emerge through the discharge of the varactor diode,
the transmission starts to recover.
In Fig.~\ref{fig3}(a), the state denoted ``EIT on'' (``EIT off'') 
corresponds to the state in which the trapped mode and the radiative mode 
are coupled (decoupled).

The dynamic switching between EIT on and EIT off states
enables the storage of electromagnetic waves by the following procedure.
First, the metamaterial is prepared in an EIT on state,
in which the radiative and trapped modes are coupled.
If the spectrum of the incident pulse is in the transparent window,
the pulse transmits through the metamaterial with the transit time
or group delay.
The delay corresponds to the time period when
the electromagnetic energy received in the radiative 
mode is transferred into the trapped mode and returns to
the radiative mode through coupling.
If during the propagation the metamaterial is changed
into an EIT off state, where the two resonant modes are decoupled,
some of the energy is captured in the trapped mode.
Then, this energy is released by returning the metamaterial to an
EIT on state to reintroduce the coupling.

Before the demonstration of pulse storage,
we measured the transmission of a pulse through the metamaterial
in an EIT on state to estimate the delay of the pulse
due to slow propagation in the metamaterial.
We prepared a $35$-$\U{ns}$ pulse with a carrier frequency of $1.440\,\U{GHz}$.
The dotted line and the solid line in Fig.~\ref{fig3}(b)
represent the power of the pulse transmitted through the waveguide without and
with the metamaterial, respectively, for $V=-0.4\,\U{V}$.
Compared with the original pulse,
a fraction of the transmitted pulse is delayed by $3$ to $4\,\U{ns}$,
which corresponds to the transit time for some components of the
input pulse within the transparent window.
(The component outside of the transparent window is blocked 
by the metamaterial.)
In other words, 
the power in the gray region is temporally stored in the trapped modes.
The energy in the gray region can be captured in the trapped
modes by switching the metamaterial from the EIT on state
to the EIT off state when the input pulse is turned off.

To demonstrate the storage and retrieval of electromagnetic waves, 
we modulate the bias voltage with a width of 
$\tau=10\,\U{ns}$, as shown in Fig.~\ref{fig3}(a)
for an input microwave pulse represented by the dotted line in
Fig.~\ref{fig3}(b),
which is redisplayed as a dotted line in Fig.~\ref{fig3}(c).
In order to hold the energy corresponding to
 the gray region in Fig.~\ref{fig3}(b),
we adjust the timing so that the original pulse is turned off at $t=0$.
Figure \ref{fig3}(c) shows the transmitted power as a solid line.
In comparison with the unmodulated case shown in Fig.~\ref{fig3}(b),
the delayed component (the gray region) is not observed
because the energy is stored in the trapped modes, which are
decoupled from the radiative modes in the EIT off state.
The stored energy starts to be radiated from the radiative mode,
when the coupling between the trapped modes and the radiative modes
is restored and the transparent window begins to open.
By comparing the height of the retrieved pulse at $t=15\,\U{ns}$
with the height of the original pulse,
the efficiency of retrieval is estimated to be about 10\%.
Figure \ref{fig3}(d) shows the magnified waveforms of retrieved pulses for 
various $\tau$.
The magnitudes of the retrieved pulses
are reduced with increasing $\tau$,
because the stored energy decays in the trapped mode
during the EIT off state.
The height of the retrieved pulses
decays exponentially with a time constant of
$6.3\,\U{ns}$, which corresponds to the lifetime of the trapped mode.

We compare retrieved pulses propagating in the 
backward and forward directions
for $n$-layered metamaterials ($n=1, 2, 3$)
by measuring the waveforms of the reflected and transmitted waves.
Figure \ref{fig3}(e) shows
retrieved pulses for forward (backward) waves on the
left (right) side. For $n=1$,
we observe almost the same amplitudes of retrieved signals
propagating in opposite directions
because the energy is radiated from a single emitter.
With increasing $n$,
the retrieved pulses become stronger for forward propagation
but diminish for backward propagation.
This indicates that the oscillation in each trapped mode
during the storage process ``inherits'' the relative phase 
of the oscillating electric field of the incident waves at
each location,
and the retrieved pulse reproduces the phase distributions 
of the incident waves.

In this Rapid Communication, we have presented a method
of dynamically modulating the properties of 
a varactor-loaded EIT metamaterial, whose characteristics can be controlled 
by an external voltage source. We have achieved 
the storage and retrieval of electromagnetic waves.
In our experiment, even though we used only three elements of the metamaterial
to avoid technical problems such as transmission reduction caused by
Ohmic loss in the varactor diodes, 
we still retrieved 10\% of the original pulse height after a storage
time of $15\,\U{ns}$.
Decay in the trapped mode, which limits the storage time,
could be compensated for by introducing an amplification process. \cite{Dong2010}

We introduced a varactor diode on one side of the element,
and the resonant frequency of the trapped mode varied,
depending on the bias voltage.
Therefore, the oscillation frequency of the trapped mode during the storage
period was different from that of the incident waves.
The retrieved pulse is frequency chirped
because energy is released from the frequency-shifting trapped mode
through the radiative mode
during the transition time $t\sub{T}$.
In order to avoid this frequency chirp,
we must reduce $t\sub{T}$ by improving the bias circuitry
or change both capacitances simultaneously, as was done in the simulation.

We have also shown that the traveling pulses can 
be stored and recovered without losing their phase information.  
These results suggest that 
if we increase the number of layers,
we could potentially store the whole of an input pulse with arbitrary
temporal shape. 
It is also possible to store waves with arbitrary transverse modes 
and any polarizations by appropriately distributing the elements.
We believe that the present approach can be applied even to 
optical regions by replacing the varactor diodes
by other nonlinear elements such as a nonlinear optical crystal
substrate, whose refractive index can be changed by auxiliary light.

The present research was supported by Grants-in-Aid
for Scientific Research No. 22109004 and No. 22560041.

\bibliographystyle{apsrev4-1}

\clearpage

\end{document}